\documentclass[journal=nalefd,manuscript=letter,layout=twocolumn]{achemso}
\setkeys{acs}{maxauthors = 0}

\usepackage[version=3]{mhchem} 
\usepackage{graphicx}
\usepackage{dcolumn}
\usepackage{bm}
\usepackage[colorlinks,linkcolor=blue,anchorcolor=blue,citecolor=blue,urlcolor=blue]{hyperref}
\usepackage[mathlines]{lineno}
\usepackage{amssymb}
\usepackage{amsmath}
\usepackage{natbib}
\usepackage{soul,color}
\usepackage{xcolor,colortbl}
\usepackage{multirow}
\usepackage{color}%

\makeatletter
\renewcommand*{\@fnsymbol}[1]{\ensuremath{\ifcase#1\or \dagger\or *\or \ddagger\or
\mathsection\or \mathparagraph\or \|\or **\or \dagger\dagger \or
\ddagger\ddagger \else\@ctrerr\fi}} \makeatother

\title{Microscopic Mechanism of the Helix-to-Layer \\
       Transformation in Elemental Group VI Solids}

\author{Dan~Liu}
\affiliation{Physics and Astronomy Department,
             Michigan State University,
             East Lansing, Michigan 48824, USA}

\author{Xianqing Lin}
\affiliation{Physics and Astronomy Department,
             Michigan State University,
             East Lansing, Michigan 48824, USA}
\alsoaffiliation{College of Science,
             Zhejiang University of Technology,
             Hangzhou 310023, China}

\author{David Tom\'{a}nek}
\email
            {tomanek@pa.msu.edu}%
\affiliation{Physics and Astronomy Department,
             Michigan State University,
             East Lansing, Michigan 48824, USA}

\date{\today} 

\abbreviations{DFT, GW, VASP, 1D, 2D}


\begin{document}


\begin{abstract}
We study the conversion of bulk Se and Te, consisting of
intertwined $a$ helices, to structurally very dissimilar,
atomically thin two-dimensional (2D) layers of these elements. Our
{\em ab initio} calculations reveal that previously unknown and
unusually stable $\delta$ and $\eta$ 2D allotropes may form in an
intriguing multi-step process that involves a concerted motion of
many atoms at dislocation defects. We identify such a complex
reaction path involving zipper-like motion of such dislocations
that initiate structural changes. With low activation barriers
${\lesssim}0.3$~eV along the optimum path, the conversion process
may occur at moderate temperatures. We find all one-dimensional
(1D) and 2D chalcogen structures to be semiconducting.
%
\end{abstract}

After much attention has been devoted to graphene, a 2D allotrope
of group IV elemental carbon, scientific interest turned to
semiconducting 2D allotropes of group V elements
P~\cite{{Li2014},{DT229}} and As~\cite{{Osters12},{Pumera17}}.
Recent observation of 2D allotropes of group VI elements Se and
Te~\cite{{ZhangPRL2017},{Chen2017},{Peide-Se17}} came as a
surprise, since -- unlike group IV and V elemental solids -- the
bulk structure of Se and Te is not layered, but consists of
helical chains of covalently bonded atoms packed in a hexagonal
array. For lack of well-defined layers, 2D Se and Te can not be
obtained by mechanical exfoliation used in group IV and V systems.
Chalcogens are known for a large number of stable allotropes and
oxidation states~\cite{Greenwood97}. The latter fact had been
identified as the key factor behind the stability of specific 2D
allotropes of Se and Te~\cite{ZhangPRL2017}. Still, the strong
dissimilarity between the bulk structure containing weakly
interacting, intertwined $a$ helices and covalently bonded,
atomically thin layers raises the intriguing question about the
microscopic mechanism behind the transformation from quasi-1D to
2D structures, which has not been addressed yet.

\begin{figure}[t]
\includegraphics[width=1.0\columnwidth]{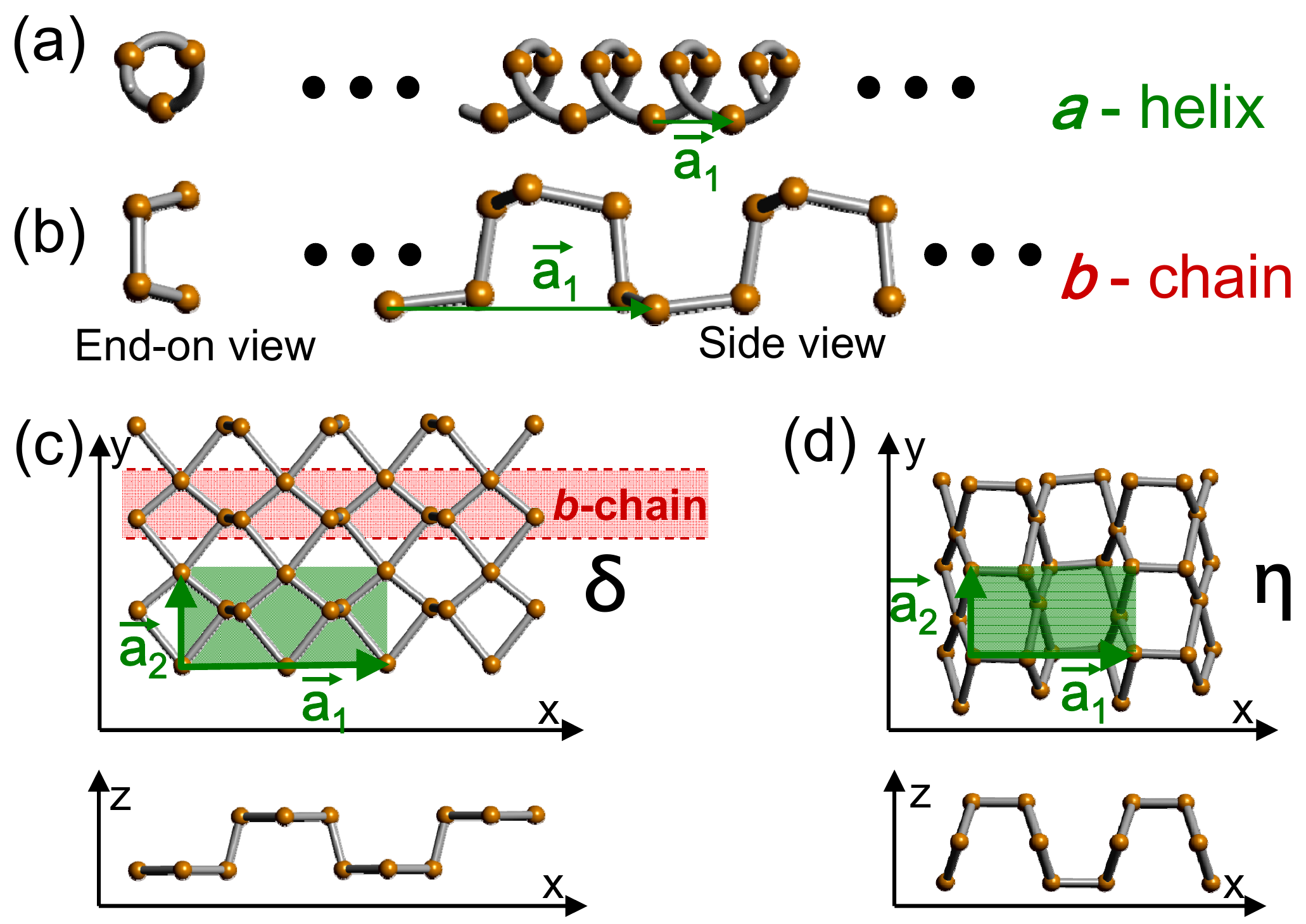}
\caption{(Color online) Stable 1D and 2D structural allotropes of
Se and Te. 1D structures of the (a) $a$ helix and (b) $b$ chain
and their 2D counterparts, the (c) $\delta$ and (d) $\eta$
allotrope. The $\delta$ allotrope is a covalently bonded 2D
assembly of $b$ chains. The unit cells of the 2D structures are
highlighted by the transparent green areas in (c) and (d).
\label{fig1}}
\end{figure}
\begin{figure}[t]
\includegraphics[width=1.0\columnwidth]{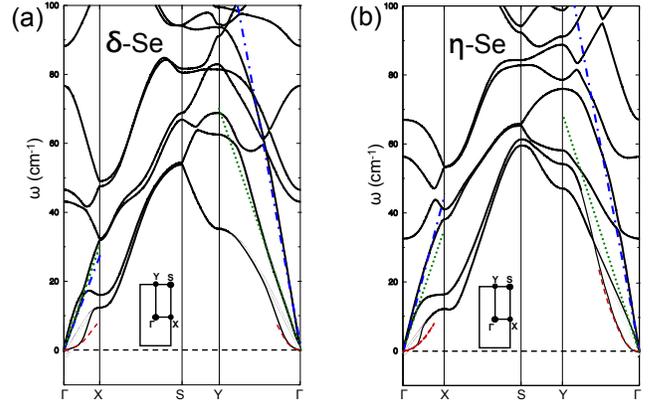}
\caption{
{(Color online) Phonon spectra of (a) $\delta$-Se and %
(b) $\eta$-Se calculated using the DFT-LDA energy functional. The
Brillouin zones and high-symmetry points are shown schematically
in the insets. Continuum elasticity results for long-wavelength
longitudinal acoustic modes are shown by the blue dash-dotted
lines, for transverse acoustic modes by the dotted green lines,
and for flexural modes by the dashed red parabolas.} %
\label{fig2}}
\end{figure}

Here we present results of \textit{ab initio} calculations, which
help to identify the intermediate steps of the observed transition
from $a$ helices in the native bulk structure to atomically thin
layers of elemental Se and Te~\cite{ZhangPRL2017}. Our results
unveil the energetics and the intermediate steps encountered
during this structural transition. We have discovered an
intriguing mechanism that converts an $a$ helix
to a more stable, previously unknown $b$ chain
by moving a point-dislocation connecting these two structures. In
a zipper-like motion, the $b$ chain may reconnect to a previously
unknown 2D $\delta$ structure, which is unusually stable, similar
to the related $\eta$ structure. The structural change from the
$a$ helix to the 2D $\delta$ allotrope is mildly exothermic with
$-0.17$~eV/atom for Se
and $-0.23$~eV/atom for Te.
The low number of structural constraints allows the helical
structure to exploit many degrees of freedom and thus to lower the
activation barriers along the reaction path to ${\lesssim}0.3$~eV,
indicating that the transition may occur at moderate temperatures.
Our GW quasiparticle calculations of the electronic structure
indicate that all quasi-1D and 2D chalcogen allotropes are
semiconducting.

\section*{Results}

\subsection*{Formation of 2D monolayers of Se and Te from 1D $a$ helices}

\begin{figure*}[t]
\includegraphics[width=1.5\columnwidth]{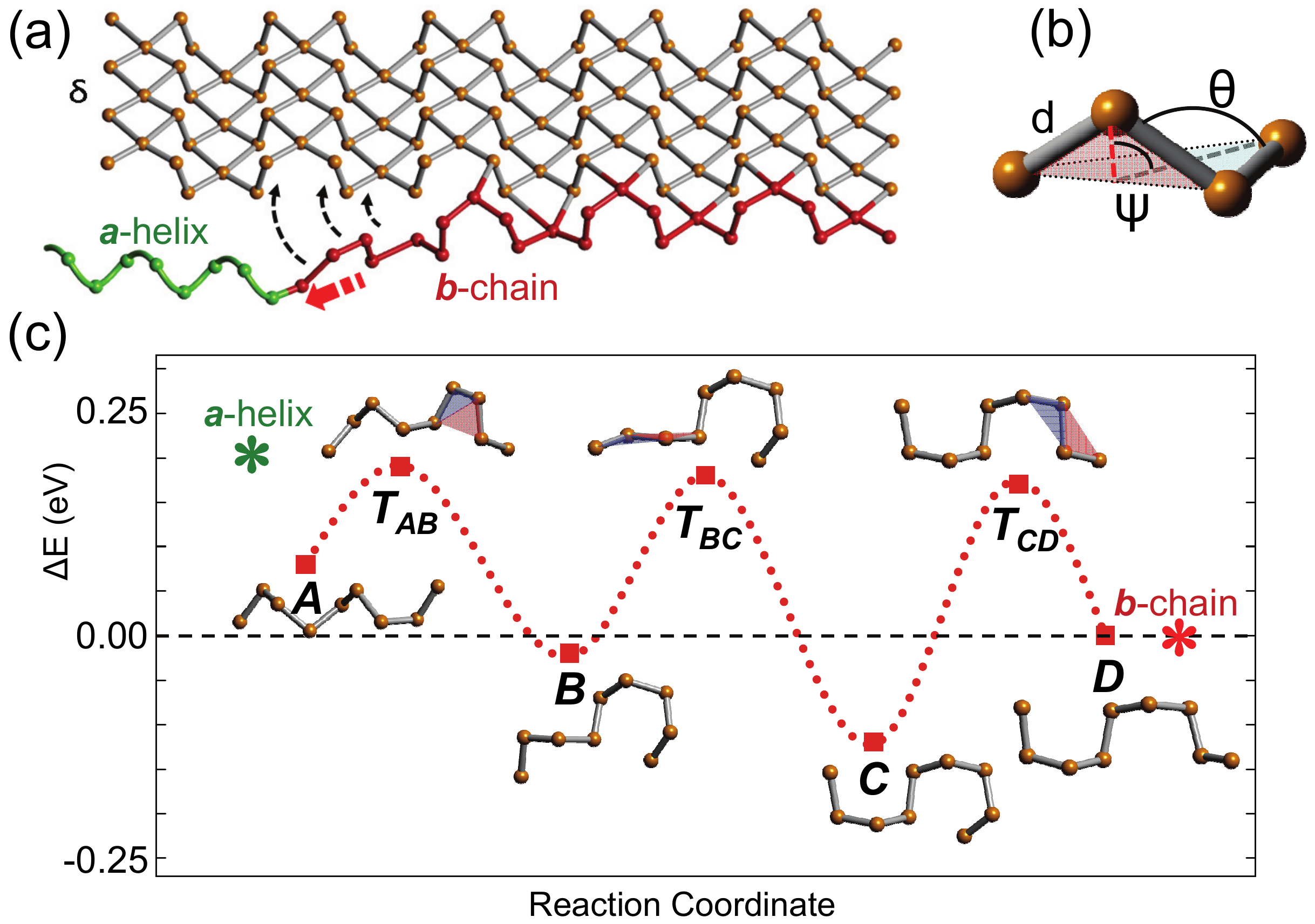}
\caption{(Color online) %
(a) Schematic growth mechanism of the energetically stable 2D
$\delta$ structure by zipper-like attachment of the $b$ chain,
which is being formed locally locally at a defect in the native
$a$ helix and propagates by dislocation motion. %
(b) Bond length $d$, bond angle $\theta$ and dihedral angle $\psi$
used to characterize chalcogen structures. %
(c) DFT-LDA based energy differences ${\Delta}E$ encountered
during the stepwise conversion from 1D $a$-Se to $b$-Se as a
function of the reaction coordinate. The system is represented by
a finite Se$_{9}$H$_2$ chain, passivated by hydrogen at both ends,
and the total energy is given with respect to the final state. The
dotted line is guide to the eye.
The energy of a $9$-atom long segment of the defect-free infinite
$a$-Se and $b$-Se chains is indicated by asterisks. Ball-and-stick
models show stable Se$_{9}$H$_2$ geometries, labeled by $A$-$D$,
and the transition states $T$. Location of the unusually small
dihedral angle in the transition states is indicated by shaded
triangles. \label{fig3}}
\end{figure*}

Understanding the observed 1D to 2D transformation is an
unprecedented challenge due to the large and constantly changing
number of degrees of freedom that are actively involved in
lowering the activation barriers between intermediate states. In
this complex system, the use of common techniques such as the
nudged elastic band %
model becomes a futile endeavor. Restricting the system's freedom
invariably increases the activation barriers, incorrectly
suggesting that the transformation should not occur under
laboratory conditions. We chose a different approach that will be
discussed in the following.

\begin{table*}[t]
\caption{%
\label{table1}%
Cohesive energy $E_{coh}$ of various Se and Te allotropes in
[eV/atom] units, obtained using DFT-LDA and DFT-PBE calculations.%
}%
\centering %
\scalebox{1.0} %
{\begin{tabular}{lcccccccc} %
\hline \hline
   \textrm{} %
 & \textrm{} %
 & \textrm{\textit{a}-helix} %
 & \textrm{\textit{b}-chain} %
 & \textrm{$\alpha$}
 & \textrm{$\beta$}
 & \textrm{$\gamma$}
 & \textrm{$\delta$}
 & \textrm{$\eta$} \\
\hline%
\multirow{2}{*}{\large Se}%
& {LDA} %
& {3.677} %
& {3.700} %
& {3.795} %
& {3.823} %
& {3.569} %
& {3.843} %
& {3.854} \\
& {PBE} %
& {3.307} %
& {3.327} %
& {3.198} %
& {3.302} %
& {2.953} %
& {3.357} %
& {3.355} \\
\hline%
\multirow{2}{*}{\large Te}
& {LDA} %
& {3.209} %
& {3.235} %
& {3.479} %
& {3.434} %
& {3.323} %
& {3.443} %
& {3.451} \\
& {PBE} %
& {2.856} %
& {2.871} %
& {2.904} %
& {2.933} %
& {2.716} %
& {2.940} %
& {2.943} \\
\hline \hline
\end{tabular}}
\label{table1}
\end{table*}

Recently observed 2D Se and Te
structures~\cite{{ZhangPRL2017},{Chen2017},{Peide-Se17},{Huang2017}}
have been formed by initially evaporating the bulk substances. It
is likely that the vapor contained primarily short segments of $a$
helices, shown in Fig.~\ref{fig1}(a), which constitute the bulk
structure. Consequently, we will consider the $a$ helix as the
initial structure in the transformation process to 2D structures.
We discovered a previously unknown, atomically thin and unusually
stable 2D structure of Se and Te, which we call the $\delta$
structure, by artificially compressing a 2D assembly of the native
$a$-Se helices in the direction normal to the 2D layer. The 2D
$\delta$ allotrope, shown in Fig.~\ref{fig1}(c), emerged
after the pressure was released. %
%
{The specific conditions for this deformation process are
specified in the Supporting Information.} %
Another previously unknown and stable allotrope, labeled $\eta$,
is depicted in Fig.~\ref{fig1}(d).
%
{It is related to the $\delta$ structure by a series of
reflections, discussed in the Supporting Information, while
keeping the bond lengths and bond angles constant throughout
the structure.} %
%
{The space group of the $\delta$ structure is $C_{2v}^4$ in the
Sch\"{o}nflies notation and its group number is \#28. The space
group of the $\eta$ structure is $D_{2}^2$ in the Sch\"{o}nflies
notation and its group number is \#17. Both groups have only 4
symmetry operations.}
Numerical results for the cohesive energies of all known Se and Te
allotropes are summarized in Table~\ref{table1}.

%

{We studied the stability of the new phases by determining their
elastic response and their phonon spectra. Since the 3D elastic
modulus tensor is not defined for a truly 2D system, which does
not naturally form layered 3D compounds, we have determined the
components of the 2D elastic tensor defined earlier~\cite{DT255}.
For the $\delta$-phase of Se, we find %
$c_{11}=4.97$~N/m, %
$c_{22}=20.02$~N/m, %
$c_{66}=5.92$~N/m, %
$D({\Gamma-X})=0.33$~eV, $D({\Gamma-Y})=1.15$~eV. %
For the $\eta$-phase of Se, we find %
$c_{11}=11.25$~N/m, %
$c_{22}=22.71$~N/m, $c_{66}=7.09$~N/m, $D({\Gamma-X})=0.39$~eV,
$D({\Gamma-Y})=1.02$~eV. %
Among others, these elastic constants allow a more accurate
representation of low-frequency acoustic modes in the vibrational
band structure of the 2D structures, which we present in
Fig.~\ref{fig2}. Due to the similarity of the phonon spectra, we
expect the zero-point motion to not to play an important role in
the cohesive energy. We find a zero-point energy of $23$~meV/atom
for $\beta$-Se and $24$~meV/atom for $\delta$-Se and $\eta$-Se,
with energy differences of ${\lesssim}0.1$~meV/atom between the
different phases.
} %

We have also confirmed the dynamic stability of the 1D and 2D
structures by performing canonical molecular dynamics (MD)
simulations at elevated temperatures. Results of $4$~ps long runs
for the $b$ chain at $300$~K and $500$~K, and of a $2$~ps run for
$\delta$-Se at $300$~K are shown as videos in the Supporting
Information.

In the following, we will first address structures of elemental Se
and refer discussion of Te structures for later. %
{Our DFT-LDA and DFT-PBE results suggest that $\delta$-Se and
$\eta$-Se are energetically near-degenerate. The previously
introduced~\cite{{ZhangPRL2017},{Chen2017}} %
$\beta$-Se allotrope is less stable than $\delta$-Se by
$20$~meV/atom (LDA) and $55$~meV/atom (PBE) and thus the least
stable of the three. Still, in view of the relatively small energy
differences and structural similarities, we expect that all these
structural allotropes, and possibly even others, may be
formed under synthesis conditions at elevated temperatures.} %
In the following, we will focus on the energetically stable
$\delta$ structure and its microscopic formation mechanism
starting from the native $a$ helix structure.

Inspecting the equilibrium structure of $\delta$-Se in
Fig.~\ref{fig1}(c), we found that it can be viewed as a 2D
assembly of 1D chains, which we call $b$ chains. We found the
previously unknown $b$ chain, shown in Fig.~\ref{fig1}(b), to be a
stable allotrope of Se, even more stable than the 1D $a$ helix by
$23$~meV/atom (LDA) and $20$~meV/atom (PBE). The $b$ chain may be
attached laterally to a semi-infinite $\delta$-Se layer in a
zipper-like motion depicted in Fig.~\ref{fig3}(a). Owing to the
multi-valent behavior of the chalcogens~\cite{ZhangPRL2017}, this
is an activation-free exothermic process that releases
$143$~meV/atom (LDA) and $30$~meV/atom (PBE).

Assuming that $b$-Se chains, which are entropically favored at
high temperatures, are indeed present during the formation of
$\delta$-Se, then the only task remaining to understand the entire
conversion path from $a$-Se to $\delta$-Se is locating an
energetically favorable pathway for the transformation from the
$a$ helix to the $b$ chain. The most plausible transformation
begins by connecting the semi-infinite $a$ helix and a $b$ chain
end-to-end by a covalent bond, as seen in Fig.~\ref{fig3}(a). The
$a$-$b$ connection is a dislocation defect or a 0D domain wall
that may propagate along the 1D chain, as indicated by the broken
arrow in Fig.~\ref{fig3}(a) and in a schematic movie in the
Supporting Information. For Se, the step-wise $a$-to-$b$
conversion is exothermic and requires only a finite activation
energy to be discussed below. Individual processes within the
entire $a$-to-$b$-to-$\delta$ transformation may occur
concurrently, as shown in Fig.~\ref{fig3}(a).

We noted that observed stable allotropes of Se and Te all share
structural commonalities. As defined in Fig.~\ref{fig3}(b), these
include the bond length $d(\rm{Se}){\approx}2.38$~{\AA} and
$d(\rm{Te}){\approx}2.84$~{\AA}, the bond angle
$\theta{\approx}100^{\circ}-130^{\circ}$, and the dihedral angle
${\psi}{\approx}80^{\circ}-100^{\circ}$ found experimentally in
bulk structures%
~\cite{{Brown96},{Cherin67},{Burbank1951},{Adenis1989},{Jamieson1965}}.
The step-wise dislocation motion corresponds to a series of $a$-Se
to $b$-Se structural changes at the dislocation, which were
studied in finite chain segments containing $12$ Se atoms,
passivated by hydrogen at both ends. Relaxing the finite segment
of $a$-Se yielded structure $A$ and relaxing the finite $b$-Se
segment resulted in structure $D$. Interestingly, the infinite $a$
and $b$ chains as well as their finite counterparts $A$ and $D$
displayed very similar structural characteristics as the bulk
structures. Later on, we found out that shorter, $9$-atom segments
shown in Fig.~\ref{fig3}(c), are sufficient to visualize and
understand the step-wise transformations in the 1D structure.

We discovered that the relative stability of the infinite $a$-Se
and $b$-Se chains, as well as that of the optimized finite
segments, can be rationalized in terms of strain originating in
the deviation from the optimum bond length $d=2.36$~{\AA}, bond
angle ${\theta}{\approx}106^\circ$ and dihedral angle
${\psi}{\approx}83^\circ$, defined in Fig.~\ref{fig3}(b). Most
robust of these parameters is the bond length, which is close to
its optimum value in all optimized structures. Whereas %
${\theta}=102^\circ$ and ${\psi}=100^\circ$ are constant
throughout the unit cell of $a$-Se, two thirds of the $b$-Se
unit cell display %
${\theta}=105^\circ$ and ${\psi}=83^\circ$, and the rest is
characterized by %
${\theta}=101^\circ$ and ${\psi}=100^\circ$. %
The closer proximity of $b$-Se to the optimum angles ${\theta}$
and ${\psi}$ is reflected in its higher stability by $23$~meV/atom
(LDA) and $20$~meV/atom (PBE) with respect to $a$-Se.




We found that the entire $A$ to $D$ transformation can be
accomplished by a sequence of bond rotations or reflection while
maintaining the optimum bond lengths $d$ and bond angles $\theta$
throughout the structure. The transformation steps only involved
changes in one dihedral angle $\psi$ at a time, as shown in
Fig.~\ref{fig3}(c), which required a typical activation energy of
${\lesssim}300$~meV. Our DFT-LDA energies for the process
described in Fig.~\ref{fig3}(c) differ from DFT-PBE results by
${\lesssim}30$~meV and from van-der-Waals corrected DFT-optB86b
results by ${\lesssim}25$~meV for the entire structure. The
similarity in cohesive energies obtained using LDA and PBE is also
seen in Table~\ref{table1}. We identified two locally stable
structures, labeled $B$ and $C$, along the $A$ to $D$ trajectory.
The locally stable states $A$, $B$, $C$ and $D$ all displayed
near-optimum values of $d$, $\theta$ and $\psi$ throughout the
structure. The contiguous trajectory in configurational space
contains unstable transition states $T_{AB}$ between $A$ and $B$,
$T_{BC}$ between $B$ and $C$, and $T_{CD}$ between $C$ and $D$. We
traced back the lower stability of the transition states to one of
the dihedral angles being near zero, far from its optimum value.
We verified that all transition states $T$ were unstable in the
sense that perturbing the $T_{N,N+1}$ structure in whichever way
and following up with microcanonical MD calculations or conjugate
gradient (CG) optimization always lead to optimum $N$ or $N+1$
geometries and to no other structure. The relative energy and the
structure of these states are depicted in Fig.~\ref{fig3}(c). More
details and MD simulations of the entire $A$ to $D$ transformation
are presented in the Supporting Information. We expect the
postulated transition process to be just one of many similar
transformations in the system that may occur with potentially even
lower activation barriers.

We should remember that the sequence and energetics of $A$ to $D$
transformations, identified in the finite chain segment, may
differ in detail from the corresponding process at a dislocation
defect connecting infinite $a$ and $b$ chains, since the
free-standing finite structure has fewer constraints than the
infinite structure. As seen in Fig.~\ref{fig3}(c), the net energy
gain from the infinite $a$ to the $b$ structure is higher than the
energy gain in the finite segment changing from the $A$ to the $D$
structure.  As a matter of fact, we should not place too much
emphasis on the relative stability of finite $A$, $B$, $C$ and $D$
structures, but rather realize that the activation barriers for
step-wise structural changes are similar in finite and infinite
structures. The energetics and structure of the free-standing
finite chain segment will change when connected to a semi-infinite
$a$ chain at the one and a semi-infinite $b$ chain at the other
end. Whatever differences in the relative stability of the
intermediate states in the infinite or finite segment connecting
$a$ and $b$ chains, none will stop the attachment of the $b$ chain
to $\delta$-Se that occurs at a significant net energy gain of
$143$~eV/atom, thus driving the exothermic reaction forward.


\begin{figure}[t]
\includegraphics[width=1.0\columnwidth]{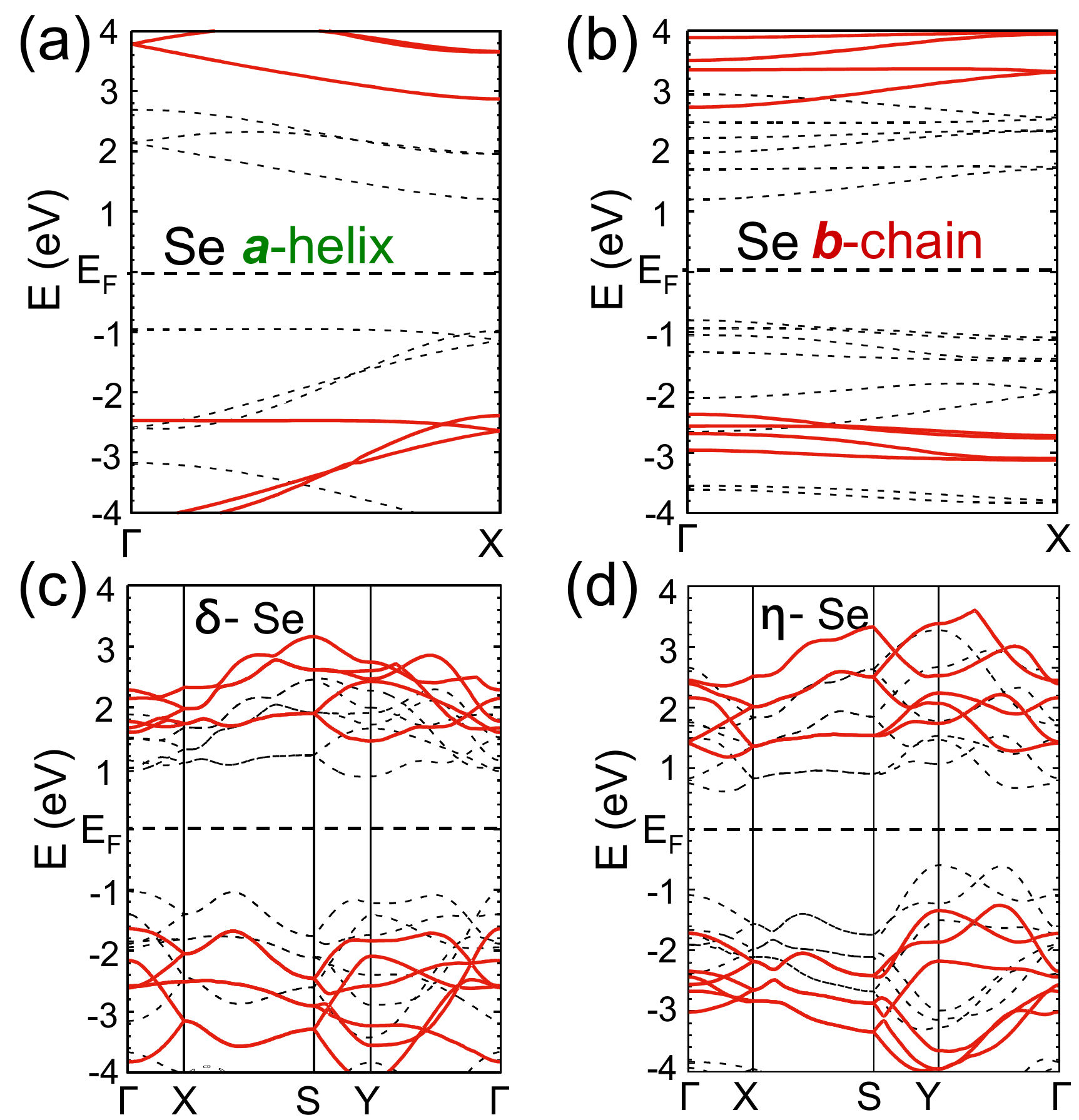}
\caption{(Color online) Electronic band structure of an isolated
(a) $a$-Se helix, (b) $b$-Se chain, isolated (c) $\delta$-Se and
(d) $\eta$-Se monolayers. GW results, shown by solid red lines,
are compared to LDA results, shown by the black dashed lines.
\label{fig4}}
\end{figure}
\begin{figure}[h]
\includegraphics[width=1.0\columnwidth]{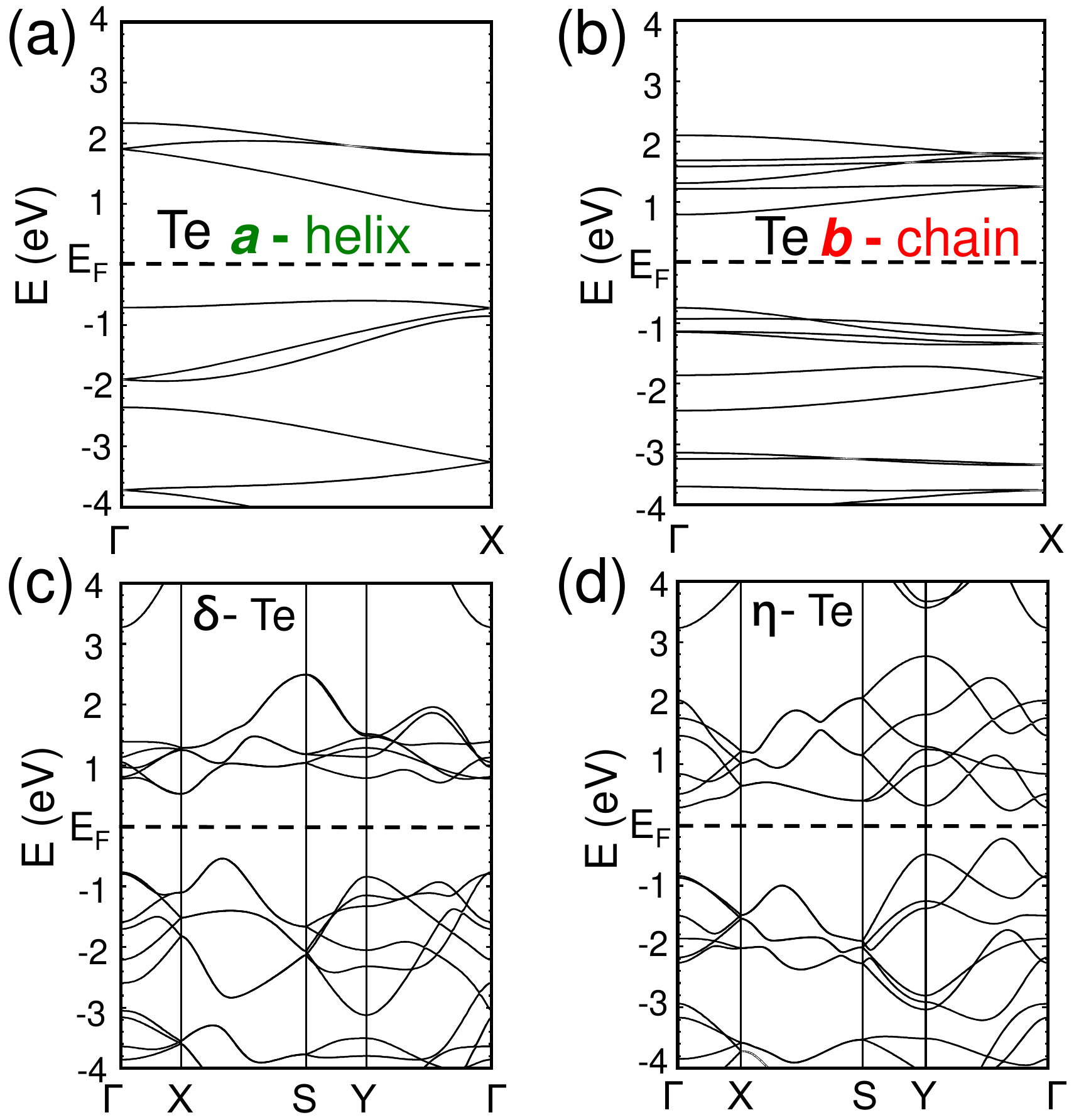}
\caption{(Color online) Electronic band structure of an isolated
(a) $a$-Te helix, (b) $b$-Te chain, isolated  (c) $\delta$-Te and
(d) $\eta$-Te monolayers obtained using DFT-LDA. %
\label{fig5}}
\end{figure}

Since Te shares the same group VI with Se in the periodic table,
we expect the chemical behavior and bonding in the respective
elemental solids to be very similar. We found the formation
mechanism of the $\delta$ allotrope from native $a$ helices via
the $b$ chains, discussed above for Se, to be viable for Te as
well, with small differences in reaction energies. According to
Table~\ref{table1}, the $a$-Te to $b$-Te conversion is exothermic,
releasing $26$~meV/atom (LDA) and $15$~meV/atom(PBE). Attaching
the $b$ chain laterally to a semi-infinite $\delta$-Te layer is
also exothermic, releasing $208$~meV/atom (LDA) and $69$~meV/atom
(PBE). As mentioned earlier, also the geometries of the initial,
final and intermediate states encountered during the $a$-Te to
$b$-Te transformation are similar, the main difference being the
Te-Te bond length, which is larger than the value in Se
allotropes. Most important, also activation energies of
${\approx}0.3$~eV are similar in Te and Se.

\subsection*{Electronic structure of 1D and 2D Se and Te allotropes }

The electronic band structure of the different 1D and 2D
allotropes of Se is shown in Fig.~\ref{fig4}. The GW results,
shown by the solid red lines, are considered a valid counterpart
to experimental observation. The LDA results, shown as a matter of
reference by the dashed black lines,
underestimate the fundamental band gap $E_g$ significantly. %
Among 1D structures, %
$a$-Se in Fig.~\ref{fig4}(a) has an $E_g=5.3$~eV
wide direct gap at $X$, and the %
$b$-Se in Fig.~\ref{fig4}(b) has an $E_g=5.1$~eV %
wide direct gap at $\Gamma$. %
Our GW band gap for $a$-Se compares favorably with a recently
reported $E_g=5.46$~eV value~\cite{andharia2017quasi} and a
smaller $E_g=3.00$~eV value~\cite{Tuttle17} obtained using a
different approach. %
$\delta$-Se in Fig.~\ref{fig4}(c) has an $E_g=3.1$~eV %
wide %
indirect gap, and %
$\eta$-Se in Fig.~\ref{fig4}(d) has an $E_g=2.4$~eV
wide indirect gap. In general, we see that the band gaps in 1D
structures are significantly larger than in the 2D layers.


The electronic band structure of different 1D and 2D allotropes of
Te is displayed in Fig.~\ref{fig5}. We find the trends and main
results for the different Te allotropes to be consistent with
those for Se, in particular the band gaps in 1D structures to be
much larger than in 2D structures. As seen in Fig.~\ref{fig5}(a),
the band gap of the isolated Te $a$ helix is $1.4$~eV wide and
indirect. The band gap of the isolated Te $b$ chain, on the other
hand, is direct at $\Gamma$ and $1.5$~eV wide, as seen in
Fig.~\ref{fig5}(b). The band structure of 2D $\delta$-Te, shown in
Fig.~\ref{fig5}(c), has an indirect, $0.9$~eV wide band gap.
Results for the 2D allotrope $\eta$-Te, shown in
Fig.~\ref{fig5}(d), indicate a direct, $0.3$~eV wide band gap
between $\Gamma$ and $Y$. These numerical results indicate that
the LDA-based band gaps in Te are roughly one third of the LDA
values found in Se. As mentioned earlier, while DFT-LDA
underestimates band gaps, it still provides useful insight into
trends in the electronic structure.

\section*{Discussion}

Reported experimental results for 2D chalcogen allotropes include
a thin layer of $a$-Se helices on silicon~\cite{Peide-Se17}, a
monolayer of $a$-Te helices on graphene~\cite{Huang2017}, a
substrate-free thin layer of $a$-Te helices~\cite{Peide-Te17}, and
$\beta$-Te, a covalently bonded 2D assembly of $a$
helices~\cite{{ZhangPRL2017},{Chen2017}}. A valid question to ask
is, why the more stable 2D $\delta$ allotrope and the 1D $b$ chain
have not been observed. Since interest in group VI elemental 2D
structures took off only very recently, it is quite possible that
optimum conditions for the synthesis of the proposed allotropes
have not been found yet. Established information about 3D
chalcogen allotropes consisting of interacting $a$ helices
provides only a limited insight into how structures may grow on a
2D substrate. There, the substrate-chalcogen interaction may play
a significant role, such as providing extra stabilization of the
more reactive $b$ chains over the $a$ helices.

Without question, the substrate plays a significant role
facilitating the 1D to 2D transformation in chalcogen structures.
Even a weak adsorbate-substrate interaction will confine
condensing chalcogen structures in the 2D space adjacent to the
substrate, thus significantly increasing the coalescence rate. To
decouple the intrinsic chalcogen reaction energetics from
substrate-chalcogen interaction on a particular substrate, we
performed all our calculations in vacuum. Specific substrates can
be selected that may change the relative stability order in
adsorbed chalcogen structures in comparison to such structures in
vacuum.

As mentioned before, locating a transformation path in
configurational space between very dissimilar structures $a$ and
$\delta$, with activation barriers not exceeding
${\approx}0.3$~eV, is a nontrivial task. So far, state-of-the-art
global structural searching techniques were unable to locate such
a path or other structures that were very different from the
native $a$ helices~\cite{ZhangPRL2017}. We feel that for the time
being, understanding the physical origin of strong bonds in terms
of $d$, $\theta$, $\psi$ and locating a pathway along which only
the least energy sensitive structural parameter is modified is a
more promising approach to understanding the reaction energy. We
located such a path from $A$ to $D$ that involves only a sequence
of changes in $\psi$ within a $9$-atom segment of the chain.

Maybe the most important lesson to learn was that -- at least in
the class of structures discussed here -- releasing structural
constraints and increasing the number of degrees of freedom may
significantly lower the activation barriers for structural
transformations. Even though the initially considered artificial
compression of a 2D assembly of $a$ helices with $3$ atoms per
unit cell to a completely flat structure with $6$ degrees of
freedom per cell did eventually yield the stable $\delta$
allotrope, the energy invested was unphysically high. Allowing for
concerted atomic motion in a $9$-atom segment with $27$ degrees of
freedom lowered the activation barriers significantly. In a
related scenario of structural phase transitions in
monochalcogenindes, artificial spatial constraints~\cite{FeiPRL16}
were also found to significantly affect the energy barriers and
thus the critical temperature~\cite{MehboudiPRL16}. The same
behavior can be expected for a wide range of systems undergoing
structural phase changes.

\section*{Summary and Conclusions}

In summary, based on DFT calculations, we have uncovered the
microscopic mechanism of the recently observed structural
transition in elemental chalcogens Se and Te from their native
bulk structure consisting of $a$ helices to atomically thin 2D
layers. We found that the $a$ helices convert to more stable,
previously unknown $b$ chains in a multi-step process that
involves a point-dislocation motion along the helix. In a
zipper-like motion, the $b$ chain reconnects to a related,
previously unknown and unusually stable 2D $\delta$ structure of
Se and Te.
The 1D $a$ helix to 2D $\delta$ conversion is mildly exothermic
with $-0.17$~eV/atom for Se
and $-0.23$~eV/atom for Te.
The high structural flexibility allows the helix to exploit many
degrees of freedom and thus significantly lower the activation
barriers along the complex reaction path to ${\lesssim}0.3$~eV,
indicating that the conversion may occur at moderate temperatures.
%
{In view of the similar stability of the structurally related
$\beta$, $\delta$ and $\eta$ structures, we expect that all these
and maybe even other allotropes should be formed at elevated
temperatures. }%
We found all 1D and 2D chalcogen structures to be semiconducting.

\section*{Computational Techniques}
Our calculations of the stability, equilibrium structure, the
pathway and dynamics of structural transformations have been
performed using density functional theory (DFT) as implemented in
the {\textsc{SIESTA}}~\cite{SIESTA} and
{\textsc{VASP}}~\cite{VASP,VASPPAW} codes. Periodic boundary
conditions have been used throughout the study, with monolayers
represented by a periodic array of slabs separated by a 30~{\AA}
thick vacuum region. We compared results using both the Local
Density Approximation (LDA)~\cite{{Ceperley1980},{Perdew81}} and
the Perdew-Burke-Ernzerhof (PBE)~\cite{PBE} exchange-correlation
functionals, since LDA typically overbinds and PBE underbinds. We
also checked the importance of van der Waals corrections to the
total energy by using the optB86b exchange-correlation
functional~\cite{{optB86b1},{optB86b2}} for selected structures.
The {\textsc{SIESTA}} calculations used norm-conserving
Troullier-Martins pseudopotentials~\cite{Troullier91}, a
double-$\zeta$ basis including polarization orbitals, and a mesh
cutoff energy of $180$~Ry to determine the self-consistent charge
density, which provided us with a precision in total energy of
${\lesssim}2$~meV/atom. The {\textsc VASP} calculations were
performed using the projector augmented wave (PAW)
method~\cite{VASPPAW} and $500$~eV as energy cutoff. The
reciprocal space has been sampled by a fine
grid~\cite{Monkhorst-Pack76} of $10{\times}10$~$k$-points in the
2D Brillouin zones (BZ) of the primitive unit cells of the
$\delta$ and $\eta$ structures containing $6$ atoms each, and
$10$~$k$-points in the BZ of 1D $a$ and $b$ chains with $3$- and
$6$-atom unit cells, respectively. Geometries have been optimized
using the conjugate gradient (CG) method~\cite{CGmethod}, until
none of the residual Hellmann-Feynman forces exceeded
$10^{-2}$~eV/{\AA}. Microcanonical and canonical MD calculations
were performed using $1$~fs time steps. Electronic structure has
been calculated using the GW quasiparticle approach~\cite{GW86} as
implemented in the {\textsc{BerkeleyGW}} package~\cite{BerkeleyGW}
interfaced with {\textsc{QuantumEspresso}}~\cite{QuantumEspresso}.
In a periodic arrangement, 1D and 2D structures were separated by
$17$~{\AA} wide vacuum regions. The Brillouin zone of quasi-1D
structures was sampled by $36{\times}1{\times}1$ k-points and that
of quasi-2D structures by $6{\times}14{\times}1$ k-points. We used
10~Ry as energy cutoff for the plane wave expansion of the
dielectric matrix. The quasiparticle energies have been determined
by considering the lowest 220 unoccupied conduction bands and
accounting for all higher-lying bands using the modified
static-remainder approximation~\cite{Deslippe13}.

\begin{suppinfo}
Detailed information regarding the dynamical stability and
{structural transformations in 1D and 2D Se allotropes. Discussed
are the transformation from the 1D $a$ helix to the 2D $\delta$
allotrope by artificial confinement, microscopic transformation
from the $\delta$ to the $\eta$ structure, and the microscopic
transformation mechanism from the $a$ helix to the $b$ chain. }%
Also provided are video files of MD simulations of a 1D $b$ chain,
a 2D $\delta$-Se allotrope, and the transformation from the 1D
$a$ helix to the $b$ chain. \\
\end{suppinfo}
\quad\par

{\noindent\bf Author Information}\\

{\noindent\bf Corresponding Author}\\
$^*$E-mail: {\tt tomanek@pa.msu.edu}

{\noindent\bf Notes}\\
The authors declare no competing financial interest.

\begin{acknowledgement}
D.L. and D.T. acknowledge financial support by the NSF/AFOSR EFRI
2-DARE grant number EFMA-1433459. Computational resources have
been provided by the Michigan State University High Performance
Computing Center. X.L. acknowledges support by the China
Scholarship Council.
\end{acknowledgement}

%

\providecommand{\latin}[1]{#1} \makeatletter \providecommand{\doi}
  {\begingroup\let\do\@makeother\dospecials
  \catcode`\{=1 \catcode`\}=2 \doi@aux}
\providecommand{\doi@aux}[1]{\endgroup\texttt{#1}} \makeatother
\providecommand*\mcitethebibliography{\thebibliography} \csname
@ifundefined\endcsname{endmcitethebibliography}
  {\let\endmcitethebibliography\endthebibliography}{}

\clearpage
\begin{figure*}[t]
\includegraphics{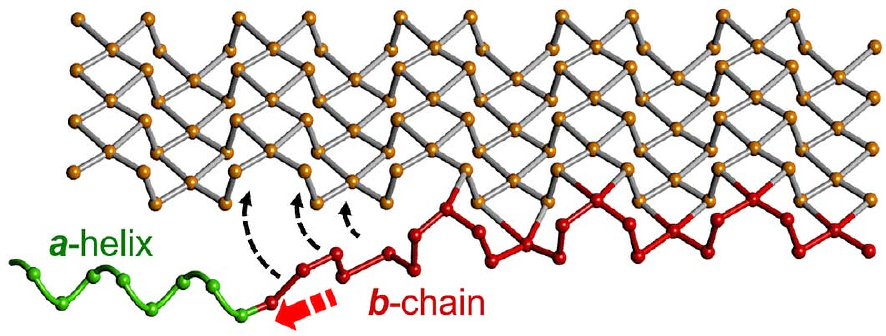}
\end{figure*}

\end{document}